\documentclass[manuscript,screen,nonacm]{acmart}
\usepackage{soul}
\AtBeginDocument{%
  \providecommand\BibTeX{{%
    \normalfont B\kern-0.5em{\scshape i\kern-0.25em b}\kern-0.8em\TeX}}}

\setcopyright{none}

\usepackage{hyperref}
\usepackage[hyphenbreaks]{breakurl}

\begin{document}

\title{Contextual Integrity of A Virtual (Reality) Classroom}


\author{Karoline Brehm}
  \email{brehm@yorku.ca}
  \orcid{https://orcid.org/0009-0007-2701-8621}
  \affiliation{
      \institution{Bauhaus-Universität Weimar}
      \country{Germany}
  }
  \affiliation{%
      \institution{York University}
      \country{Canada}
  }

  \author{Yan Shvartzshnaider}
  \orcid{0000-0001-5954-916X}
  \email{yansh@yorku.ca}
  \affiliation{%
      \institution{York University}
      \country{Canada}
  }

  \author{David Goedicke}
  \email{dg536@cornell.edu}
  \orcid{0000-0002-4837-893X}
  \affiliation{%
      \institution{Cornell Tech}
      \country{USA}
  }

  \settopmatter{printacmref=false}

\begin{abstract}
The multicontextual nature of immersive VR makes it difficult to ensure contextual integrity of VR-generated information flows using existing privacy design and policy mechanisms. In this position paper, we call on the HCI community to do away with lengthy disclosures and permissions models and move towards embracing privacy mechanisms rooted in Contextual Integrity theory.
\end{abstract}





\maketitle

\section{Motivation}

The advent of Virtual Reality (VR) technologies fostered new modes of learning, connecting and socialising in virtual worlds that can defy the laws of physics. Bounded solely by VR creators' imaginations, users can seamlessly teleport themselves to a magical fairyland, a doctor's office, a lecture hall and other virtual spaces. Behind the VR ``magic'', however, lie a multitude of sensors that support core functions of VR devices, all the while generating vast amounts of data about VR users (e.g. body and eye movements, voice, facial expressions, biometrics, behaviour) and their surroundings (e.g. room layouts, objects, bystanders)~\cite{XRSIPrivacySafety2022}. Moreover, recent studies~\cite{nairExploringUnprecedentedPrivacy2022, hellerWatchingAndroidsDream2021} have shown that this data can reveal additional information about physiology, cognition and identity of a VR user.  As the VR technology advances, it is important to evaluate the potential privacy risks and harms \cite{abrahamImplicationsXRPrivacy2022,adamsEthicsEmergingStory2018}, before its mainstream adoption in contexts such as education~\cite{asieduMetaNextBig2022,yoshimuraStudyClassMeetings2021,combsMetaFundsVirtual2022},  workplace~\cite{pidelCollaborationVirtualAugmented2020} and healthcare\cite{pillaiImpactVirtualReality2019}.

In this position paper, we explore privacy challenges posed by VR technology in the education context based on a virtual classroom case study. Hailed as ``the New Era of Learning Experience''~\cite{MetaverseVirtualClassroom2022} by its creators, virtual classrooms take full advantage of the VR platform to augment the traditional learning experience, with a range of interactive activities and tools for collaboration. In these co-created virtual spaces, students and teachers can participate from anywhere in the world, seamlessly moving between different classrooms and activities.

While virtual classrooms mimic the real-world experience, the information collection practices behind VR extend beyond the normative expectation of a traditional classroom. The ostensible enhancement of the learning experience relies on a constant supply of VR users' data. Reminiscent of existing digital platforms, the extensive data collection, however, opens up the new education setting to a range of privacy and security threats~\cite{cohneyVirtualClassroomsReal2021,carterWhatAreRisks2021}. This is complicated by the fact that some data practices are necessary to enable core VR functionality. Furthermore, in VR platforms today informed consent mechanisms prevail that fail to adequately informing users and lack effective mechanisms of outing out. As VR users navigate a multitude of contexts, a different approach is needed to address privacy challenges in both policy and design which we articulate in this paper. 
\section{The Multicontextual Policy Challenge}

A student attending a lecture in VR from home is acting both as a student in the classroom and as a member of a household. This means that the student operates in multiple social contexts. These contexts are governed by different privacy norms~\cite{nissenbaumPrivacyContextTechnology2010}: data processing in the context of the home may clash with what is considered appropriate in the virtual classroom. For example, optical and depth sensors on the students' VR headsets map the environment. While this is necessary for navigating in the virtual classroom, the same sensors can capture other people in the vicinity~\cite{mcgillExtendedRealityXR2021}, such as family members or roommates.
A similar issue arose with the introduction of video conferences and online educational platforms at universities, which caused privacy violations due to a mismatch of students' expectations and data practices of the platforms~\cite{cohneyVirtualClassroomsReal2021}.

Existing privacy-enhancing approaches in VR are predominately based on the notions of transparency and ``informed consent''~\cite{EthicalConsiderationsExtended2021, XRSIPrivacySafety2022,xrassociationChapterDesigningImmersive2022}. Suggestions include ``contextual privacy communications'' at the time of data processing~\cite{XRSIPrivacySafety2022}, visualisation of privacy risks and customisable privacy settings ~\cite{kimVirtualRealityData2022}.  In doing so, the burden of understanding the risks and preventing harm in VR falls disproportionately on the user. This further contributes to ``[widening] the gap between the users’ understanding of the types of data collected, the implications of their consent and the actual implications''~\cite{kimVirtualRealityData2022}. 
 
The multicontextual nature of VR applications calls for privacy-enhancing mechanisms that incorporate privacy norms in multiple overlapping contexts. In our work, we draw on the theory of Contextual Integrity (CI)~\cite{nissenbaumPrivacyContextTechnology2010} to address these issues. Specifically, how can we ensure that privacy norms are respected in VR-based platforms and what CI-based methodologies could be employed by policymakers and designers?
\section{VR Privacy-Preserving Design Challenges}

One of the main challenges in designing a privacy-preserving VR experience is maintaining contextual integrity of users' data. Specifically, this entails developing mechanisms to ensure that information flows in accordance with users' expectations and established societal norms. A suitable solution would need to differentiate between a multitude of nested contexts. For example, in a VR classroom, a conversation between fellow students is subject to different privacy norms than a meeting with the academic supervisor or a student counsellor. Privacy protections would also need to account for  different modes of interaction and be robust to accommodate the privacy norms associated with future VR adaptations. 

In search of answers to these challenges, the research community is  building tools to help users make \emph{meaningful} privacy-related choices. Examples of such efforts include: designing a VR-native interface for privacy settings, in lieu of a traditional text-based menu~\cite{yaoVirtualEquipmentSystem2021}; providing users with an option to engage in an ``incognito mode" to stop the flow of information to other VR users~\cite{yaoVirtualEquipmentSystem2021}; and informing the user about privacy risks through  a role-playing privacy tutorial that visually reenacts the information flow prescribed by a privacy policy~\cite{limMineYourselfRoleplaying2022}.

These privacy-preserving approaches, however helpful, still build on privacy models that embody  notions of information secrecy, control, personal anonymity and informed consent, despite their well-documented shortcomings~\cite{cateFailureFairInformation2006, schaubDesignSpaceEffective2015}. Here, too, with close coordination between all the relevant stakeholders of the VR ecosystem\footnote{We envision collaboration between VR users, designers, system engineers, developers, policymakers and others.}, the theory of CI can help inform the design of new privacy protection mechanisms based on the notion of appropriate information flows in accordance with enforced privacy norms. The CI-based approach would rely on tools for identifying information flows generated in the system~\cite{shvartzshnaiderVACCINEUsingContextual2019}  and prescribed by the policy~\cite{shvartzshnaiderGoingAppropriateFlow2019}, capturing users' expectations~\cite{shvartzshnaiderLearningPrivacyExpectations2016, apthorpeDiscoveringSmartHome2018}, and devising effective mechanisms of detecting deviation from contextual privacy norms in a VR environment.

\section{Acknowledgements}
This project has been funded by the Office of the Privacy Commissioner of Canada (OPC); the views expressed herein are those of the author(s) and do not necessarily reflect those of the OPC.

\bibliographystyle{ACM-Reference-Format}
\bibliography{references.bib}


\begin{thebibliography}{26}


\ifx \showCODEN    \undefined \def \showCODEN     #1{\unskip}     \fi
\ifx \showDOI      \undefined \def \showDOI       #1{#1}\fi
\ifx \showISBNx    \undefined \def \showISBNx     #1{\unskip}     \fi
\ifx \showISBNxiii \undefined \def \showISBNxiii  #1{\unskip}     \fi
\ifx \showISSN     \undefined \def \showISSN      #1{\unskip}     \fi
\ifx \showLCCN     \undefined \def \showLCCN      #1{\unskip}     \fi
\ifx \shownote     \undefined \def \shownote      #1{#1}          \fi
\ifx \showarticletitle \undefined \def \showarticletitle #1{#1}   \fi
\ifx \showURL      \undefined \def \showURL       {\relax}        \fi
\providecommand\bibfield[2]{#2}
\providecommand\bibinfo[2]{#2}
\providecommand\natexlab[1]{#1}
\providecommand\showeprint[2][]{arXiv:#2}

\bibitem[Abraham et~al\mbox{.}(2022)]%
        {abrahamImplicationsXRPrivacy2022}
\bibfield{author}{\bibinfo{person}{Melvin Abraham}, \bibinfo{person}{Pejman
  Saeghe}, \bibinfo{person}{Mark McGill}, {and} \bibinfo{person}{Mohamed
  Khamis}.} \bibinfo{year}{2022}\natexlab{}.
\newblock \showarticletitle{Implications of {XR} on {Privacy}, {Security} and
  {Behaviour}: {Insights} from {Experts}}. In \bibinfo{booktitle}{\emph{{ACM}
  {International} {Conference} {Proceeding} {Series}}}.
  \bibinfo{publisher}{Association for Computing Machinery}.
\newblock
\showISBNx{978-1-4503-9699-8}
\urldef\tempurl%
\url{https://doi.org/10.1145/3546155.3546691}
\showDOI{\tempurl}


\bibitem[Adams et~al\mbox{.}(2018)]%
        {adamsEthicsEmergingStory2018}
\bibfield{author}{\bibinfo{person}{Devon Adams}, \bibinfo{person}{Alseny Bah},
  \bibinfo{person}{Catherine Barwulor}, \bibinfo{person}{Nureli Musaby},
  \bibinfo{person}{Kadeem Pitkin}, {and} \bibinfo{person}{Elissa~M. Redmiles}.}
  \bibinfo{year}{2018}\natexlab{}.
\newblock \showarticletitle{Ethics {Emerging}: the {Story} of {Privacy} and
  {Security} {Perceptions} in {Virtual} {Reality}}. In
  \bibinfo{booktitle}{\emph{Fourteenth {Symposium} on {Usable} {Privacy} and
  {Security} ({SOUPS} 2018)}}. \bibinfo{publisher}{USENIX Association},
  \bibinfo{address}{Baltimore, MD}, \bibinfo{pages}{427--442}.
\newblock
\showISBNx{978-1-939133-10-6}
\urldef\tempurl%
\url{https://www.usenix.org/conference/soups2018/presentation/adams}
\showURL{%
\tempurl}


\bibitem[Apthorpe et~al\mbox{.}(2018)]%
        {apthorpeDiscoveringSmartHome2018}
\bibfield{author}{\bibinfo{person}{Noah Apthorpe}, \bibinfo{person}{Yan
  Shvartzshnaider}, \bibinfo{person}{Arunesh Mathur}, \bibinfo{person}{Dillon
  Reisman}, {and} \bibinfo{person}{Nick Feamster}.}
  \bibinfo{year}{2018}\natexlab{}.
\newblock \showarticletitle{Discovering {Smart} {Home} {Internet} of {Things}
  {Privacy} {Norms} {Using} {Contextual} {Integrity}}.
\newblock \bibinfo{journal}{\emph{Proceedings of the ACM on Interactive,
  Mobile, Wearable and Ubiquitous Technologies}} \bibinfo{volume}{2},
  \bibinfo{number}{2} (\bibinfo{date}{July} \bibinfo{year}{2018}),
  \bibinfo{pages}{59:1--59:23}.
\newblock
\urldef\tempurl%
\url{https://doi.org/10.1145/3214262}
\showDOI{\tempurl}


\bibitem[Asiedu(2022)]%
        {asieduMetaNextBig2022}
\bibfield{author}{\bibinfo{person}{Kwasi~Gyamfi Asiedu}.}
  \bibinfo{year}{2022}\natexlab{}.
\newblock \bibinfo{title}{Meta’s next big bet: {The} ‘metaversity'}.
\newblock
\newblock
\urldef\tempurl%
\url{https://www.protocol.com/enterprise/metaverse-in-education-morehouse-meta}
\showURL{%
\tempurl}
\newblock
\shownote{Publisher: protocol}.


\bibitem[Carter and Egliston(2021)]%
        {carterWhatAreRisks2021}
\bibfield{author}{\bibinfo{person}{Marcus Carter} {and} \bibinfo{person}{Ben
  Egliston}.} \bibinfo{year}{2021}\natexlab{}.
\newblock \showarticletitle{What are the risks of {Virtual} {Reality} data?
  {Learning} {Analytics}, {Algorithmic} {Bias} and a {Fantasy} of {Perfect}
  {Data}}.
\newblock \bibinfo{journal}{\emph{New Media \& Society}} \bibinfo{volume}{0},
  \bibinfo{number}{0} (\bibinfo{year}{2021}),
  \bibinfo{pages}{14614448211012794}.
\newblock
\urldef\tempurl%
\url{https://doi.org/10.1177/14614448211012794}
\showDOI{\tempurl}
\newblock
\shownote{\_eprint: https://doi.org/10.1177/14614448211012794}.


\bibitem[Cate(2006)]%
        {cateFailureFairInformation2006}
\bibfield{author}{\bibinfo{person}{Fred~H. Cate}.}
  \bibinfo{year}{2006}\natexlab{}.
\newblock \bibinfo{title}{The {Failure} of {Fair} {Information} {Practice}
  {Principles}}.
\newblock
\newblock
\urldef\tempurl%
\url{https://papers.ssrn.com/abstract=1156972}
\showURL{%
\tempurl}


\bibitem[Cohney et~al\mbox{.}(2021)]%
        {cohneyVirtualClassroomsReal2021}
\bibfield{author}{\bibinfo{person}{Shaanan Cohney}, \bibinfo{person}{Ross
  Teixeira}, \bibinfo{person}{Anne Kohlbrenner}, \bibinfo{person}{Arvind
  Narayanan}, \bibinfo{person}{Mihir Kshirsagar}, \bibinfo{person}{Yan
  Shvartzshnaider}, {and} \bibinfo{person}{Madelyn Sanfilippo}.}
  \bibinfo{year}{2021}\natexlab{}.
\newblock \showarticletitle{Virtual {Classrooms} and {Real} {Harms}: {Remote}
  {Learning} at {U}.{S}. {Universities}}. In
  \bibinfo{booktitle}{\emph{Seventeenth {Symposium} on {Usable} {Privacy} and
  {Security} ({SOUPS} 2021)}}. \bibinfo{publisher}{USENIX Association},
  \bibinfo{pages}{653--674}.
\newblock
\showISBNx{978-1-939133-25-0}
\urldef\tempurl%
\url{https://www.usenix.org/conference/soups2021/presentation/cohney}
\showURL{%
\tempurl}


\bibitem[Combs(2022)]%
        {combsMetaFundsVirtual2022}
\bibfield{author}{\bibinfo{person}{Veronica Combs}.}
  \bibinfo{year}{2022}\natexlab{}.
\newblock \bibinfo{title}{Meta funds virtual classrooms at 7 universities
  complete with {VR} headsets}.
\newblock
\newblock
\urldef\tempurl%
\url{https://www.techrepublic.com/article/meta-funds-virtual-classrooms-at-7-universities-complete-with-vr-headsets/}
\showURL{%
\tempurl}


\bibitem[{Edverse}(2022)]%
        {MetaverseVirtualClassroom2022}
\bibfield{author}{\bibinfo{person}{{Edverse}}.}
  \bibinfo{year}{2022}\natexlab{}.
\newblock \bibinfo{title}{Metaverse {Virtual} {Classroom}: {The} {New} {Era}
  {Learning} {Experience}}.
\newblock
\newblock
\urldef\tempurl%
\url{https://www.edverse.com/blog/virtual-classrooms/}
\showURL{%
\tempurl}


\bibitem[Heller(2021)]%
        {hellerWatchingAndroidsDream2021}
\bibfield{author}{\bibinfo{person}{Brittan Heller}.}
  \bibinfo{year}{2021}\natexlab{}.
\newblock \showarticletitle{Watching {Androids} {Dream} of {Electric} {Sheep}:
  {Immersive} {Technology}, {Biometric} {Psychography}, and the {Law}}.
\newblock \bibinfo{journal}{\emph{Vanderbilt Journal of Entertainment and
  Technology Law}} \bibinfo{volume}{23}, \bibinfo{number}{1}
  (\bibinfo{year}{2021}).
\newblock
\urldef\tempurl%
\url{https://scholarship.law.vanderbilt.edu/jetlaw/vol23/iss1/1}
\showURL{%
\tempurl}


\bibitem[{IEEE Standards Association}(2021)]%
        {EthicalConsiderationsExtended2021}
\bibfield{author}{\bibinfo{person}{{IEEE Standards Association}}.}
  \bibinfo{year}{2021}\natexlab{}.
\newblock \bibinfo{title}{Ethical {Considerations} of {Extended} {Reality}
  ({XR})}.
\newblock
\newblock
\urldef\tempurl%
\url{https://beyondstandards.ieee.org/ethical-considerations-of-extended-reality-xr/}
\showURL{%
\tempurl}


\bibitem[Kim(2022)]%
        {kimVirtualRealityData2022}
\bibfield{editor}{\bibinfo{person}{Yeji Kim}} (Ed.).
  \bibinfo{year}{2022}\natexlab{}.
\newblock \showarticletitle{Virtual {Reality} {Data} and {Its} {Privacy}
  {Regulatory} {Challenges}: {A} {Call} to {Move} {Beyond} {Text}-{Based}
  {Informed} {Consent}}.
\newblock \bibinfo{journal}{\emph{California Law Review}}
  \bibinfo{volume}{110}, \bibinfo{number}{1} (\bibinfo{year}{2022}),
  \bibinfo{pages}{225}.
\newblock
\urldef\tempurl%
\url{https://doi.org/10.15779/Z380Z70X6P}
\showDOI{\tempurl}
\newblock
\shownote{Publisher: California Law Review;}.


\bibitem[Lim et~al\mbox{.}(2022)]%
        {limMineYourselfRoleplaying2022}
\bibfield{author}{\bibinfo{person}{Junsu Lim}, \bibinfo{person}{Hyeonggeun
  Yun}, \bibinfo{person}{Auejin Ham}, {and} \bibinfo{person}{Sunjun Kim}.}
  \bibinfo{year}{2022}\natexlab{}.
\newblock \showarticletitle{Mine {Yourself}!: {A} {Role}-playing {Privacy}
  {Tutorial} in {Virtual} {Reality} {Environment}}. In
  \bibinfo{booktitle}{\emph{Extended {Abstracts} of the 2022 {CHI} {Conference}
  on {Human} {Factors} in {Computing} {Systems}}} \emph{(\bibinfo{series}{{CHI}
  {EA} '22})}. \bibinfo{publisher}{Association for Computing Machinery},
  \bibinfo{address}{New York, NY, USA}, \bibinfo{pages}{1--7}.
\newblock
\showISBNx{978-1-4503-9156-6}
\urldef\tempurl%
\url{https://doi.org/10.1145/3491101.3519773}
\showDOI{\tempurl}


\bibitem[McGill(2021)]%
        {mcgillExtendedRealityXR2021}
\bibfield{author}{\bibinfo{person}{Mark McGill}.}
  \bibinfo{year}{2021}\natexlab{}.
\newblock \bibinfo{booktitle}{\emph{Extended {Reality} ({XR}) and the {Erosion}
  of {Anonymity} and {Privacy}}}.
\newblock \bibinfo{type}{{T}echnical {R}eport}. \bibinfo{institution}{The IEEE
  Global Initiative on Ethics of Extended Reality}.
\newblock
\urldef\tempurl%
\url{https://standards.ieee.org/wp-content/uploads/import/governance/iccom/extended-reality-anonymity-privacy.pdf}
\showURL{%
\tempurl}


\bibitem[Nair et~al\mbox{.}(2022)]%
        {nairExploringUnprecedentedPrivacy2022}
\bibfield{author}{\bibinfo{person}{Vivek Nair},
  \bibinfo{person}{Gonzalo~Munilla Garrido}, {and} \bibinfo{person}{Dawn
  Song}.} \bibinfo{year}{2022}\natexlab{}.
\newblock \bibinfo{title}{Exploring the {Unprecedented} {Privacy} {Risks} of
  the {Metaverse}}.
\newblock
\newblock
\urldef\tempurl%
\url{https://doi.org/10.48550/arXiv.2207.13176}
\showDOI{\tempurl}
\newblock
\shownote{arXiv:2207.13176 [cs]}.


\bibitem[Nissenbaum(2010)]%
        {nissenbaumPrivacyContextTechnology2010}
\bibfield{author}{\bibinfo{person}{Helen~Fay Nissenbaum}.}
  \bibinfo{year}{2010}\natexlab{}.
\newblock \bibinfo{booktitle}{\emph{Privacy in context: technology, policy, and
  the integrity of social life}}.
\newblock \bibinfo{publisher}{Stanford Law Books}, \bibinfo{address}{Stanford,
  Calif}.
\newblock
\showISBNx{978-0-8047-5236-7 978-0-8047-5237-4}


\bibitem[Pidel and Ackermann(2020)]%
        {pidelCollaborationVirtualAugmented2020}
\bibfield{author}{\bibinfo{person}{Catlin Pidel} {and} \bibinfo{person}{Philipp
  Ackermann}.} \bibinfo{year}{2020}\natexlab{}.
\newblock \showarticletitle{Collaboration in {Virtual} and {Augmented}
  {Reality}: {A} {Systematic} {Overview}}. In
  \bibinfo{booktitle}{\emph{Augmented {Reality}, {Virtual} {Reality}, and
  {Computer} {Graphics}}} \emph{(\bibinfo{series}{Lecture {Notes} in {Computer}
  {Science}})}, \bibfield{editor}{\bibinfo{person}{Lucio~Tommaso De~Paolis}
  {and} \bibinfo{person}{Patrick Bourdot}} (Eds.). \bibinfo{publisher}{Springer
  International Publishing}, \bibinfo{address}{Cham},
  \bibinfo{pages}{141--156}.
\newblock
\showISBNx{978-3-030-58465-8}
\urldef\tempurl%
\url{https://doi.org/10.1007/978-3-030-58465-8_10}
\showDOI{\tempurl}


\bibitem[Pillai and Mathew(2019)]%
        {pillaiImpactVirtualReality2019}
\bibfield{author}{\bibinfo{person}{Anitha~S. Pillai} {and}
  \bibinfo{person}{Prabha~Susy Mathew}.} \bibinfo{year}{2019}\natexlab{}.
\newblock \bibinfo{title}{Impact of {Virtual} {Reality} in {Healthcare}: {A}
  {Review}}.
\newblock
\newblock
\urldef\tempurl%
\url{https://doi.org/10.4018/978-1-5225-7168-1.ch002}
\showDOI{\tempurl}
\newblock
\shownote{ISBN: 9781522571681 Pages: 17-31 Publisher: IGI Global}.


\bibitem[Schaub et~al\mbox{.}(2015)]%
        {schaubDesignSpaceEffective2015}
\bibfield{author}{\bibinfo{person}{Florian Schaub}, \bibinfo{person}{Rebecca
  Balebako}, \bibinfo{person}{Adam~L. Durity}, {and}
  \bibinfo{person}{Lorrie~Faith Cranor}.} \bibinfo{year}{2015}\natexlab{}.
\newblock \showarticletitle{A design space for effective privacy notices}. In
  \bibinfo{booktitle}{\emph{Proceedings of the eleventh {USENIX} conference on
  usable privacy and security}} \emph{(\bibinfo{series}{{SOUPS} '15})}.
  \bibinfo{publisher}{USENIX Association}, \bibinfo{address}{USA},
  \bibinfo{pages}{1--17}.
\newblock
\showISBNx{978-1-931971-24-9}
\newblock
\shownote{Number of pages: 17 Place: Ottawa, Canada}.


\bibitem[Shvartzshnaider et~al\mbox{.}(2019a)]%
        {shvartzshnaiderGoingAppropriateFlow2019}
\bibfield{author}{\bibinfo{person}{Yan Shvartzshnaider}, \bibinfo{person}{Noah
  Apthorpe}, \bibinfo{person}{Nick Feamster}, {and} \bibinfo{person}{Helen
  Nissenbaum}.} \bibinfo{year}{2019}\natexlab{a}.
\newblock \showarticletitle{Going against the ({Appropriate}) {Flow}: {A}
  {Contextual} {Integrity} {Approach} to {Privacy} {Policy} {Analysis}}.
\newblock \bibinfo{journal}{\emph{Proceedings of the AAAI Conference on Human
  Computation and Crowdsourcing}}  \bibinfo{volume}{7} (\bibinfo{date}{Oct.}
  \bibinfo{year}{2019}), \bibinfo{pages}{162--170}.
\newblock
\showISSN{2769-1349}
\urldef\tempurl%
\url{https://doi.org/10.1609/hcomp.v7i1.5266}
\showDOI{\tempurl}


\bibitem[Shvartzshnaider et~al\mbox{.}(2019b)]%
        {shvartzshnaiderVACCINEUsingContextual2019}
\bibfield{author}{\bibinfo{person}{Yan Shvartzshnaider},
  \bibinfo{person}{Zvonimir Pavlinovic}, \bibinfo{person}{Ananth Balashankar},
  \bibinfo{person}{Thomas Wies}, \bibinfo{person}{Lakshminarayanan
  Subramanian}, \bibinfo{person}{Helen Nissenbaum}, {and}
  \bibinfo{person}{Prateek Mittal}.} \bibinfo{year}{2019}\natexlab{b}.
\newblock \showarticletitle{{VACCINE}: {Using} {Contextual} {Integrity} {For}
  {Data} {Leakage} {Detection}}. In \bibinfo{booktitle}{\emph{The {World}
  {Wide} {Web} {Conference}}} \emph{(\bibinfo{series}{{WWW} '19})}.
  \bibinfo{publisher}{Association for Computing Machinery},
  \bibinfo{address}{New York, NY, USA}, \bibinfo{pages}{1702--1712}.
\newblock
\showISBNx{978-1-4503-6674-8}
\urldef\tempurl%
\url{https://doi.org/10.1145/3308558.3313655}
\showDOI{\tempurl}


\bibitem[Shvartzshnaider et~al\mbox{.}(2016)]%
        {shvartzshnaiderLearningPrivacyExpectations2016}
\bibfield{author}{\bibinfo{person}{Yan Shvartzshnaider},
  \bibinfo{person}{Schrasing Tong}, \bibinfo{person}{Thomas Wies},
  \bibinfo{person}{Paula Kift}, \bibinfo{person}{Helen Nissenbaum},
  \bibinfo{person}{Lakshminarayanan Subramanian}, {and}
  \bibinfo{person}{Prateek Mittal}.} \bibinfo{year}{2016}\natexlab{}.
\newblock \showarticletitle{Learning {Privacy} {Expectations} by
  {Crowdsourcing} {Contextual} {Informational} {Norms}}.
\newblock \bibinfo{journal}{\emph{Proceedings of the AAAI Conference on Human
  Computation and Crowdsourcing}}  \bibinfo{volume}{4} (\bibinfo{date}{Sept.}
  \bibinfo{year}{2016}), \bibinfo{pages}{209--218}.
\newblock
\showISSN{2769-1349}
\urldef\tempurl%
\url{https://doi.org/10.1609/hcomp.v4i1.13271}
\showDOI{\tempurl}


\bibitem[{XR Association}(2022)]%
        {xrassociationChapterDesigningImmersive2022}
\bibfield{author}{\bibinfo{person}{{XR Association}}.}
  \bibinfo{year}{2022}\natexlab{}.
\newblock \showarticletitle{Chapter 4: {Designing} {Immersive} {Learning} for
  {Secondary} {Education}}.
\newblock In \bibinfo{booktitle}{\emph{{XR} {Assiociation} {Developers}
  {Guide}: {An} {Industry}-{Wide} {Collaboration} for {Better} {XR}}}.
  \bibinfo{publisher}{XR Association}.
\newblock


\bibitem[{XR Safety Initiative}(2022)]%
        {XRSIPrivacySafety2022}
\bibfield{author}{\bibinfo{person}{{XR Safety Initiative}}.}
  \bibinfo{year}{2022}\natexlab{}.
\newblock \bibinfo{booktitle}{\emph{The {XRSI} {Privacy} and {Safety}
  {Framework}}}.
\newblock \bibinfo{type}{{T}echnical {R}eport}. \bibinfo{institution}{XR Safety
  Initiative}.
\newblock
\urldef\tempurl%
\url{https://xrsi.org/publication/the-xrsi-privacy-framework}
\showURL{%
\tempurl}


\bibitem[Yao et~al\mbox{.}(2021)]%
        {yaoVirtualEquipmentSystem2021}
\bibfield{author}{\bibinfo{person}{Powen Yao}, \bibinfo{person}{Vangelis
  Lympouridis}, {and} \bibinfo{person}{Michael Zyda}.}
  \bibinfo{year}{2021}\natexlab{}.
\newblock \showarticletitle{Virtual equipment system: {Face} mask and voodoo
  doll for user privacy and self-expression options in virtual reality}. In
  \bibinfo{booktitle}{\emph{2021 {IEEE} conference on virtual reality and {3D}
  user interfaces abstracts and workshops ({VRW})}}. \bibinfo{publisher}{IEEE},
  \bibinfo{address}{Lisbon, Portugal}, \bibinfo{pages}{747--748}.
\newblock
\urldef\tempurl%
\url{https://doi.org/10.1109/VRW52623.2021.00256}
\showDOI{\tempurl}


\bibitem[Yoshimura and Borst(2021)]%
        {yoshimuraStudyClassMeetings2021}
\bibfield{author}{\bibinfo{person}{Andrew Yoshimura} {and}
  \bibinfo{person}{Christoph Borst}.} \bibinfo{year}{2021}\natexlab{}.
\newblock \showarticletitle{A {Study} of {Class} {Meetings} in {VR}: {Student}
  {Experiences} of {Attending} {Lectures} and of {Giving} a {Project}
  {Presentation}}.
\newblock \bibinfo{journal}{\emph{Frontiers in Virtual Reality}}
  \bibinfo{volume}{2} (\bibinfo{date}{May} \bibinfo{year}{2021}).
\newblock
\urldef\tempurl%
\url{https://doi.org/10.3389/frvir.2021.648619}
\showDOI{\tempurl}


\end{thebibliography}
\end{document}